\documentclass[aps,prd,showpacs,nofootinbib,preprintnumbers,twocolumn]{revtex4-1}
\usepackage{bm}
\usepackage{latexsym}
\usepackage{natbib}
\usepackage{url}
\usepackage{dcolumn}
\usepackage{color}
\usepackage{amsfonts,amssymb,amsmath}
\usepackage{graphicx,epsfig}
\usepackage{psfrag}
\usepackage{subfigure}
\usepackage{natbib}
\begin{document}

\title{Cosmic expansion driven by  real scalar field for different forms of potential}
\author{Murli Manohar Verma}
\email{sunilmmv@yahoo.com}

\author{Shankar Dayal Pathak}
\email{prince.pathak19@gmail.com}
\affiliation{Department of Physics, University of Lucknow, \\ Lucknow 226 007, India}
\date{May 22, 2013}

\begin{abstract}
We discuss  the expansion of the universe in the  FRLW model assuming that the source of dark energy is  either tachyonic scalar field or quintessence. The  tachyonic scalar field with exponential and power-law  potential (function of homogeneous scalar field $\phi$) both gives exponential expansion of the universe. It is found that this behaviour is not distinguishable from the quintessence with respect to these potentials.
\end{abstract}

\maketitle

\section{Introduction}
The observed accelerated expansion of the universe is understood by different dark energy models. A class of scalar fields is one of the promising candidate of dark energy \cite{i1,j1,k1}. The dark energy is treated as scalar field (tachyoinc or quintessence), among itself, the  tachyonic scalar field arising from string theory \cite{n1} (for different reasons in our context)  has been  widely used in literature \cite{l1,m1,o1}.

The relativistic Lagrangian proposed \cite{n1} for the tachyonic scalar field $\phi$ gives  the action as
\begin{eqnarray}\mathcal{A}=\int d^{4}x \sqrt{-g} \left(\frac{R}{16\pi G}- V(\phi)\sqrt{1-\partial^{i}\phi\partial_{i}\phi}\right)\label{k1}\end{eqnarray}
with tachyonic scalar Lagrangian

\begin{eqnarray}L_{tach}=-V(\phi)\sqrt{1-\partial_{i}\phi\partial^{i}\phi}\label{n1}\end{eqnarray}
and corresponding energy momentum tensor

\begin{eqnarray} T^{ik}=\frac{\partial{L}}{\partial{(\partial_{i}\phi)}}\partial^{k}\phi-g^{ik}L\label{n2}\end{eqnarray}
gives the energy density and pressure as
\begin{eqnarray} \rho=\frac{V(\phi)}{\sqrt{1-\partial_{i}\phi\partial^{i}\phi}}; \qquad P=-V(\phi)\sqrt{1-\partial_{i}\phi\partial^{i}\phi}\label{n3}\end{eqnarray} respectively. The spatially homogeneous approximation of the field leads to the following forms of the above expressions
\begin{eqnarray}\rho=\frac{V(\phi)}{\sqrt{1-\dot{\phi^{2}}}} ;\qquad P=-V(\phi)\sqrt{1-\dot{\phi^{2}}}.\label{n4}\end{eqnarray}
where an overdot denotes the derivative with respect to time (as in the rest of this paper).
The conservation equation of energy for this field is
\begin{eqnarray}\frac{\dot{\rho}_{\phi}}{\rho_{\phi}}=-3H\dot{\phi}^{2}\label{n5}\end{eqnarray}
where $H$ represents  the Hubble parameter.

The Friedmann equation for single component dominated by scalar field can be written for flat geometry is
\begin{equation}\label{n6}
  H^{2}=\frac{1}{3M_{p}^{2}}\frac{V(\phi)}{\sqrt{1-\dot{\phi}^{2}}}
\end{equation}
where $M_{p}^{2}=\frac{1}{8\pi  G}$
The equation of motion of tachyonic scalar field could be found by varying action (\ref{k1}) as

\begin{equation}\label{n6}
  \frac{\ddot{\phi}}{\dot{\phi}}+ \frac{(1-\dot{\phi}^{2})V'(\phi)}{\dot{\phi}V(\phi)} + 3H(1-\dot{\phi}^{2})  = 0
\end{equation}
here ($'$) and ($\dot{}$) stand for derivative with respect to $\phi$ and time $t$ respectively

The energy density and pressure for spatially homogeneous quintessence scalar field can be obtained from the Lagrangian
\begin{equation}\label{l1}
  L_{quin}= \frac{1}{2}\partial_{i}\phi\partial^{i}\phi -V(\phi)
\end{equation}
as
\begin{equation}\label{l2}
  \rho_{\phi}= \frac{1}{2}\dot{\phi}^{2} + V(\phi) \qquad  P_{\phi}= \frac{1}{2}\dot{\phi}^{2} + V(\phi)
\end{equation}
and equation of motion is
\begin{equation}\label{l3}
\ddot{\phi} + 3H\dot{\phi}^{2} + V'(\phi) = 0
\end{equation} with

\begin{equation}\label{k1}
  H^{2}= \frac{1}{3M_{p}^{2}}\left[\frac{1}{2}\dot{\phi}^{2} + V(\phi)\right].
\end{equation}
For acceleration we have equation of state $w< -\frac{1}{3}\Rightarrow \dot{\phi}^{2}<\frac{2}{3}$.

\section{Cosmic expansion with exponential and power law potentials}
In this section we discuss the form of scale factor for two particular potentials in two classes of scalar field (tachyonic and quintessence). The expansion of the universe is driven by exponential form as discussed below for each class.

\subsection{Tachyonic scalar field with exponential potential}

Here we assume that the form of exponential potential is
\begin{equation}\label{l4}
 V(\phi)=V_{0}\exp(-\alpha \phi)
\end{equation}
where $\alpha$ is constant. In inflation theory (slow-roll model), the slow-roll of scalar field occurs  during very small interval of time gives the exponential form of scale factor. We assume the late time accelerated expansion is probably same as early inflation. The slow-roll approximation of early inflation could be taken in context of late time accelerated expansion and we allow this approximation during long period of time.  With consideration of the potential of the form (\ref{l4}) the Hubble parameter with slow-roll assumption(i.e. $\dot{\phi}^{2}<< 1$) for present accelerated expansion era can be written as

\begin{equation}\label{l5}
  H = \frac{\dot{a(t)}}{a(t)} = \sqrt{\frac{V_{0}}{3M_{p}^{2}}}\exp \left(-\frac{1}{2}\alpha\phi\right)
\end{equation}
and the equation of motion of tachyon field with $\ddot{\phi}\simeq 0$ leads to

\begin{equation}\label{l6}
  3H\dot{\phi}  = -\frac{V^{'}(\phi)}{V(\phi)} = \alpha
\end{equation}
 equations (\ref{l5}) and (\ref{l6}) give
\begin{equation}\label{l7}
\dot{\phi}(t) = \frac{\alpha}{3\beta}\exp\left(\alpha\frac{\phi}{2}\right)
\end{equation}
where $\beta=\sqrt{\frac{M^{2}_{p}}{3V_{0}}}$. The solution of the differential equation (\ref{l7}) gives
\begin{equation}\label{l8}
 \phi(t)= -\frac{2}{\alpha}\ln\left[-\frac{\alpha^{2}}{6\beta}t + \exp(-\alpha\phi_{i}/2)\right]
\end{equation} where $\phi_{i}$ be the value of $\phi$ at time $t=t_{i}\sim 0$.
Using (\ref{l8}) in (\ref{l5}) we get expression of scale factor as function of time is

\begin{equation}\label{l9}
  \frac{a(t)}{a_{i}}= \exp\beta t\left(-\frac{\alpha^{2}}{12\beta}t + \exp(-\alpha\phi_{i}/2)\right)
\end{equation}
where $a_{i}$ is scale factor at time $t=t_{i}\sim 0$.

Consider the other exponential form of potential as
\begin{equation}
V(\phi)=V_{0}\exp \left(-\dfrac{\alpha\phi^{\gamma}}{2}\right).\label{z1}
\end{equation}

This kind of potential gives \begin{equation}
\frac{1}{\Theta}\int^{\phi}_{\phi_{i}} \phi^{1-\gamma}\exp \left(-\dfrac{\alpha\phi^{\gamma}}{2}\right)d\phi=t\label{z2}
\end{equation}
where $\Theta=\dfrac{3M^{2}_{p}\alpha\gamma}{\sqrt{3V_{0}}}$.

\begin{equation}
\Theta(2-\gamma)t= \phi^{2-\gamma}\exp \left(-\dfrac{\alpha\phi^{\gamma}}{2}\right)+\dfrac{\gamma\alpha}{2}f(\phi) \label{z3}
\end{equation}
where \begin{equation}
f(\phi)=\int^{\phi}_{\phi_{i}} \phi^{\frac{2-\gamma}{\gamma}}\exp \left(-\dfrac{\alpha\phi^{\gamma}}{2}\right) d\phi  .\label{z4}
\end{equation}

After finding $ \phi(t) $ from (\ref{z3}) and (\ref{z4}) it can be  derived the scale factor as function of time from the given expression

 \begin{equation}
\int^{a}_{a_{i}}\frac{da}{a(t)}=\sqrt{\frac{V_{0}}{3M_{p}^{2}}}\int^{t}_{0} \exp \left(-\frac{\alpha\phi^{\gamma}}{2}\right) dt  .\label{z5}
\end{equation}


\subsection{Tachyonic scalar field with power law potential}
Using the same mechanism  for the exponential potential using in the case of power law potential
\begin{equation}\label{l10}
  V(\phi)=V_{0}\phi^{n}
\end{equation}
we have the expression of $\phi(t)$ is

\begin{equation}\label{l11}
  \phi(t)= \left[\phi^{\frac{n+2}{2}}_{i}+\frac{\alpha(n+2)}{6\beta}t\right]^{\frac{2}{n+2}}
\end{equation}
where $n$ is real.
Using this expression of $\phi(t)$ in (\ref{n6}) with conditions ($\dot{\phi}^{2}<< 1$ and $\ddot{\phi}\simeq 0$) we get the expression of scale factor as function of time for power law scalar potential is

\begin{equation}\label{l12}
  \frac{a(t)}{a_{i}}=\exp\left[ \frac{3\beta^{2}}{\alpha(n+1)}\left\{-\phi^{n+1}_{i} + \left(\phi^{\frac{(n+2)}{2}}_{i} + \frac{\alpha(n+2)}{6\beta}t\right)^{\frac{2(n+1)}{(n+2)}}\right\}\right].
\end{equation}


Assume the other form of power law potential is
\begin{equation}\label{z6}
  V(\phi)=V_{0}(A \phi^{\chi} + B \phi^{\chi-1})
\end{equation}
where $A$, $B$ and $\chi$ are real constants.
Using equation of motion with slow-roll approximation for the form of potential given by (\ref{z6}) we can find $\phi(t)$ as

\begin{equation}\label{z7}
 \sqrt{\frac{3V_{0}}{M^{2}_{p}}}\int^{\phi}_{\phi_{i}} \frac{(A \phi^{\chi} + B \phi^{\chi-1})^{\frac{3}{2}}}{(A \chi \phi^{\chi-1}+ B (\chi-1) \phi^{\chi-2})} d\phi = t .\end{equation}

Scale factor of expansion can be derived from following expression
\begin{equation}
\int^{a}_{a_{i}}\frac{da}{a(t)}=\sqrt{\frac{V_{0}}{3M_{p}^{2}}}\int^{t}_{0} (A \phi^{\chi} + B \phi^{\chi-1})^\frac{1}{2} dt  .\label{z8}
\end{equation}


\section{Cosmic expansion with exponential and power law potential of quintessence scalar field}

\subsection{Quintessence scalar field with exponential potential}
The Hubble parameter
\begin{equation}\label{j1}
  H = \frac{\dot{a(t)}}{a(t)} = \sqrt{\frac{V_{0}}{3M_{p}^{2}}}\exp \left(-\frac{1}{2}\alpha\phi\right)
\end{equation}
and equation of motion of quintessence field with $\ddot{\phi}\simeq 0$ give

\begin{equation}\label{j2}
  3H\dot{\phi}  = -V^{'} = \alpha V(\phi)
\end{equation}
these two equations (\ref{j1}) and (\ref{j2}) gives
\begin{equation}\label{j3}
\dot{\phi}(t) = \eta \exp\left(-\alpha\frac{\phi}{2}\right)
\end{equation}
where
\begin{equation}\label{o1}
  \eta=\sqrt{\frac{\alpha^{2}M^{2}_{p}V_{0}}{3}}.
\end{equation}

Solution of (\ref{j3}) gives the functional form of $ \phi(t)$ as
\begin{equation}\label{j4}
 \phi(t)= \frac{2}{\alpha}\ln\left[\frac{\eta\alpha}{2}t + \exp(\alpha\phi_{i}/2)\right].
\end{equation}
Using (\ref{j4}) in (\ref{k1}) with $\dot{\phi}^{2}<< 1$, provide the scale factor as function of time is

\begin{equation}\label{j5}
  \frac{a(t)}{a_{i}}= \exp \left[\ln\left\{\frac{\eta\alpha}{2}t + \exp(\alpha\phi_{i}/2)\right\}^{\frac{2\beta}{\eta\alpha}}-\frac{\beta\phi_{i}}{\eta}\right].
\end{equation}


We use the other form of potential given by (\ref{z1}) to find $\phi(t)$ from (\ref{j1})
\begin{equation}
\frac{1}{\mu}\int^{\phi}_{\phi_{i}} \phi^{1-\gamma}\exp \left(-\dfrac{\alpha\phi^{\gamma}}{2}\right)d\phi=t \label{y1}
\end{equation}
where $\mu= -\dfrac{\alpha V_{0}M_{p}\gamma}{\sqrt{3}}$.
Scale factor of expansion for quintessence scalar field in case of this exponential form of potential is given by (\ref{z5}).


\subsection{Quintessence scalar field with power law potential}

we have expression of $\phi(t)$ is

\begin{equation}\label{j6}
  \phi(t)= \left[\phi^{\frac{4-n}{2}}_{i}-\frac{n(4-n)V_{0}}{6\beta}t\right]^{\frac{2}{4-n}}.
\end{equation}

Using this expression of $\phi(t)$  with conditions ($\dot{\phi}^{2}<< 1$) in (\ref{k1})  and condition $\ddot{\phi}\simeq 0$) in (\ref{l3}) we get following exponential form of  scale factor

\begin{equation}\label{j7}
  \frac{a}{a_{i}}=\exp \left[-\frac{3\beta^{2}}{2nV_{0}}\left\{   \phi^{\frac{4-n}{2}}_{i}  - \frac{n(4-n)V_{0}}{6\beta}t                           \right\}^{\frac{4}{4-n}}  -\phi^{2}_{i} \right].
\end{equation}




The other power law form of potential given by (\ref{z6}), we have the following expression for $\phi(t)$
\begin{equation}\label{y2}
 \sqrt{\frac{3V_{0}}{M^{2}_{p}}}\int^{\phi}_{\phi_{i}} \frac{(A \phi^{\chi} + B \phi^{\chi-1})^{\frac{1}{2}}}{(A \chi \phi^{\chi-1}+ B (\chi-1) \phi^{\chi-2})} d\phi = t.\end{equation}
From (\ref{z8}) we can get scale factor as function of time with the help of expression (\ref{y2}).

\section{Conclusion}

The mathematical analysis is  carried out, and the implications are investigated into,   for two classes  of scalar field,  namely tachyonic and quintessence with slow-roll asumptions. We consider two forms  of potential,  one is exponential function of $\phi$ and other is some power of scalar field. It has been found that the tachyonic scalar field gives exponential expansion for both potentials. This is the same  behaviour as in case of quintessence fields and so is not distinguishable from it.  The mechanism to find potential for quasi-exponential form of scale factor was  discussed  in \cite{a4} and  here we have chosen a  reverse operation for a given potential to explore  the evolution  of  scale factor as function of time.

\begin{acknowledgments}
 The authors are thankful to the University Grants Commission, New Delhi for its support through the Major Research Project vide  F. No.37-431/2009 (SR).
\end{acknowledgments}

\end{document}